# Driver Choice Compared to Controlled Diversion for a Freeway Double On-Ramp

L. C. Davis, Physics Dept., University of Michigan, Ann Arbor, MI 48109, USA

Two diversion schemes that apportion demand between two on-ramps to reduce congestion and improve throughput on a freeway are analyzed. In the first scheme, drivers choose to merge or to divert to a downstream on-ramp based on information about average travel times for the two routes: (1) merge and travel on the freeway or (2) divert and travel on a surface street with merging downstream. The flow, rate of merging at the ramps, and the travel times oscillate strongly, but irregularly, due to delayed feedback. In the second scheme, diversion is controlled by the average mainline velocities just upstream of the on-ramps. Driver choice is not involved. If the average upstream velocity on the mainline drops below a predetermined value (20 m/s) vehicles are diverted to the downstream ramp. When the average mainline velocity downstream becomes too low, diversion is no longer permitted. The resultant oscillations in this scheme are nearly periodic. The period is dominated by the response time of the mainline to interruption of merging rather than delayed feedback, which contributes only a minor component linear in the distance separating the on-ramps. In general the second scheme produces more effective congestion reduction and greater throughput. Also the travel times for on-ramp drivers are less than that obtained by drivers who attempt to minimize their own travel times (first scheme).

*E-mail address:* lloyddavis07@comcast.net.

# 1. Introduction

Schönhof *et al.* [1] have discussed how advances in inter-vehicle communication improve information flow on traffic networks. It is reasonable to assume that in the near future detailed information will be available to drivers, enabling them to make decisions to optimize their travel time. Wahle *et al.* [2] have studied decision dynamics pertaining to alternative route choices when drivers are provided travel-time data. These authors analyzed a two-route model where route choices were made on the basis of transit times. Because the information is inherently delayed, feedback caused large oscillations in the flow and number of vehicles on each route. Lee *et al.* [3] demonstrated that decisions based on the average velocity of vehicles on each route, which is more up-to-date than travel time, produced flow that was more stable.

In contrast, freeway on-ramp metering generally depends on local conditions and the response times of traffic to changes in the rate of vehicle merging. Modern control is often based on measurements of mainline occupancy (essentially density of vehicles) in the vicinity of an on-ramp [4-8]. Kerner [9-11] has criticized this approach and has proposed a method based on velocity measurement on the mainline just upstream of the merging area. In either case, the rate of merging is restricted with traffic signals to improve flow and reduce mainline congestion. Long queues can therefore develop on the on-ramp if demand is high—a problem of concern to engineers designing freeway ramp control strategy (See Xin *et al.* [12].).

In the present work, diversion of on-ramp vehicles to another on-ramp downstream via a parallel route of lower speed (such as a surface street) is analyzed. Diversion is way to reduce congestion and long queues of vehicles waiting to enter the freeway. Two options are considered: (1) on-ramp drivers decide to divert if the travel time is less for the low-speed parallel route and (2) vehicles are diverted (by signal rather than driver choice) to the downstream on-ramp if the mainline velocity near the merging area drops below a critical velocity. In this method, if the velocity measured near the downstream on-ramp drops below the critical velocity, diversion is no longer permitted. Thus this analysis of a double on-ramp scenario contains aspects of both decision dynamics and on-ramp control. It is expected that delayed feedback will play an important role.

Diverting freeway traffic around a bottleneck using local streets has been discussed by Munoz and Laval [13]. They made the simplifying assumption that travel times on surface streets are fixed and not subject to flow dynamics or congestion. The analysis of Banks [14] took into account the sensitivity of travel times to flow. In that work it was assumed that flow equilibria of the type postulated by Wardrop [15] are established by drivers who attempt to minimize their travel times. In neither paper were detailed driver-vehicle models considered. Consequently, the large oscillations predicted by Wahle *et al*. [2] were not found in those studies.

Kerner and Klenov [16] have investigated the influence of congestion at one bottleneck on the flow at an upstream on-ramp. Two empirically observed effects, induced phase transitions and the catch effect, have been demonstrated. In the present work, the interaction of on-ramps will undoubtedly influence the results. The traffic model of Kerner and Klenov, which has been extensively studied and was used in previous work on double on-ramps, will also be used in the current work. The description of the model is given in Refs. [9, 11, 16, 17] and will not be repeated in this paper. The model reproduces the various phases of traffic identified in empirical data and is computationally efficient for simulations.

The paper is organized as follows. In Sec. 2, the throughput for a double on-ramp is calculated for a given incoming flow. The maximum throughput is determined by varying the proportion of demand that merges at each ramp. In Sec. 3, vehicles approaching the upstream on-ramp can divert to the downstream on-ramp via a lower speed diversion route. Two schemes for diversion are considered: (1) drivers choose on the basis on average travel times and (2) vehicles are diverted according to average velocities on the mainline near the ramps. Conclusions are drawn in Sec. 4.

## 2. Maximizing Flow for the Double On-Ramp

As a preliminary to what follows, the optimal proportion of fixed total on-ramp demand routed to two on-ramps is determined in this section. The ramps are separated by a distance $L_{merge}$ as shown in Fig. 1.

In the example considered, the incoming flow on the mainline is 1700 vehicles/h in each of two lanes, which is less than the flow $q_{out}$ = 1810/h/lane needed to sustain a wide moving jam. The on-ramp flow is $q^{(D)}$ to the downstream ramp (D) and $q^{(U)}$ to the upstream ramp (U) with $q^{(D)} + q^{(U)}$ = 1000 vehicles/h, which is large enough to cause congestion of the on-ramp if all vehicles try to merge at the same ramp. Here $L_{merge}$ = 1000 m.

In Fig. 2a, the number of merges in an hour at U and D, as well as the total, are plotted as a function of $Q^{(U)}$, the demand in one hour at U. The corresponding demand at D is $Q(D)$ = 1000 – $Q^{(U)}$. The total number of merges is close to 1000 with a slight increase with greater $Q^{(U)}$ up to 700 and then falls off. The number of individual ramp merges closely track demand for each on-ramp.

The mainline flow just downstream of D shows a modest increase with increasing $Q^{(U)}$ in Fig 2b up to 700. Scatter in the results is due to the stochastic nature of the Kerner-Klenov model [9, 11, 17]. If all the demand is routed to just one on-ramp, congestion on the ramp sets in, limiting the number of merges to approximately 800-850 and the total flow to fewer than 4000 vehicles in one hour. Splitting the demand between two on-ramps avoids such congestion and improves overall flow. Finding larger flow with a higher proportion of the demand at U is consistent with remarks by Zhang and Recker [18] in a related context regarding metering rates for multiple on-ramps. They advised that downstream ramps should be restricted more strongly that upstream ramps.

## 3. Diversion on a Parallel Route

In this section, diversion of vehicles from the merge region of an on-ramp to a downstream merge region is considered. The diversion route is a single-lane road, such as an access road or surface street, with a lower speed limit than the freeway. The lead vehicle in the merge region of the on-ramp can be diverted to the diversion route and subsequently merge downstream, a distance of $L_{merge}$ away. The purpose of the diversion is to split the demand between two ramps to avoid congestion and improve overall flow. It can also reduce the frustration of drivers waiting to merge. No vehicles on the mainline are allowed to exit.

3.1 *Diversion based on travel time*

Two schemes for selection of vehicles to divert will be studied. In the first, detectors *Det1a* (at the beginning of the merge region U) and *Det3* (100 m downstream of D) are used to record the times vehicles pass over them (Fig. 3). Detectors *Det1* and *Det2* are not used in this scheme. The time for a vehicle to travel between the detectors is denoted by T. If the vehicle merges at the upstream ramp, it travels by path U. If the vehicle is diverted, it travels by path D. For either path, an average travel time is updated according to

$$T_{path} = \alpha T + (1-\alpha) T_{path}^{prev} \qquad (1)$$

every time a vehicle completes the route. The value chosen for *α* is a compromise between reasonable smoothing so that drivers are not confused by wild fluctuations and being as nearly current as possible. It was found that $\alpha = 0.1$ is a suitable choice. Initially, $T_{path}$ is set equal to a default value which is the route distance divided by the speed limit on that route. Also, if there are no vehicles on the diversion route (including the downstream merge region), $T_{path}^{(D)}$ is reset to the default value. In this scheme, the lead vehicle in the upstream merge region diverts when $T_{path}^{(D)} < T_{path}^{(U)}$, provided the headway to last vehicle on the diversion route is large enough. The required distance between the last vehicle and diverting vehicle is taken to be its velocity $v$ times a headway time $h_d = 1$ s plus the length of a vehicle, $\ell = 7.5$ m. That is,

$$x_{last} - x > vh_d + \ell \qquad (2)$$

Otherwise the vehicle tries to merge. It is assumed that on-ramp drivers are rational agents, *i.e.*, they always choose the best route for themselves based on a comparison of the current measured values of $T_{path}^{(D)}$ and $T_{path}^{(U)}$. These are inherently delayed by the time taken by the most recent vehicles which have completed the routes D and U.

Unlike the studies of Wahle *et al.* [2] and others, the two routes in the present analysis are inequivalent. The U route is dominated by vehicles on the mainline other than those that have merged whereas the D route has only vehicles that have been diverted. The speed limits on the two routes are also unequal with the diversion route having a lower limit. The limits are 30 m/s (108 km/h) and 22.2 m/s (80 km/h), respectively.

The results of a typical simulation are shown in Figs. 4 – 7. The average travel times for the U and D paths, $T_{path}^{(U)}$ and $T_{path}^{(D)}$, are shown as a function of time for one hour in Fig. 4. The incoming flow on the mainline is 1700 vehicles/h in each lane and the distance between merge regions is $L_{merge} = 1000$ m. The on-ramp demand is 1000 vehicles/h. Because a region of synchronous flow builds up on the freeway, the average travel time for U gradually increases while the average travel time for D fluctuates strongly, resetting to the default value when the path contains no vehicles.

In Fig. 5, the elapsed time from passing $x = -300$ m is shown as a function of the position of the 100$^{th}$ vehicle to merge at U. The velocity, which dropped sharply upon merging, is also depicted. The vehicle merged at time $t = 361$ s and at position $x = -281$ m. It took 22 s before the vehicle passed the end of the merge region at $x = 0$. Its velocity was 17 m/s at this point and it accelerated to 30 m/s over the next 600 m and 26 s. It took a total of 64 s to reach *Det3* at $x = 1100$ m, compared to the default value of 47 s. There was no congested region at D because no merges had taken place there. Thus, the congestion near U added 17 s of delay. The average travel time $T_{path}^{(U)}$ increased more and surpassed the default value of $T_{path}^{(D)} = 63$ s at $t = 463$ s. (Recall that $T_{path}$ is a weighted average of previous values of individual vehicle transit times, Eq. (1).) Once vehicles merge at D, congestion can build up and reduce the mainline velocity upstream, further increasing $T_{path}^{(U)}$.

The cumulative number of merges at each merge region is shown in Fig. 6 corresponding to the average travel times in Fig. 4. In one hour, 496 vehicles merged at U and 451 at D. Although not optimally proportioned according to the results presented in Fig. 2b, the total flow beyond D was a respectable 4014 vehicles in one hour. Likewise, the region of synchronous flow with average speed below 20 m/s was confined (in the upstream direction) to within about 3 km of U (See Fig. 7a.), which is comparable to the results for the more nearly optimal demand with $q^{(U)} > q^{(D)}$ in the double on-ramp configuration of Fig. 1. Although the feedback via the average travel times is delayed, the flow appears to be stable. A few fluctuations to low velocity occurred during the last 800 s as can be seen in Fig. 7b, but no jams were formed.

The *characteristic* oscillations in flow observed by Wahle *et al.* [2] were nearly periodic. In the present simulation, the patterns were not found to be so closely periodic. However, the average travel time for the D route shown in Fig. 4 was found to oscillate somewhat regularly as $T_{path}^{(D)}$ reset to the default value when there were no vehicles on the D path. Likewise, there are irregular gaps in merging at U as can be seen in Fig. 6. Merging at D has small gaps (small enough so it is difficult to see them in Fig. 6) roughly equal to the free-flow transit time of that route. These gaps also occur when there are no vehicles on the D path. Presumably the inequivalence of the routes in the present configuration is responsible for at least a portion of the qualitatively different behavior compared to that found by Wahle *et al.* [2]. Run-to-run variability has been found in different simulations using different random number sequences.

In Wahle *et al.* [2] and Lee *et al.* [3], the objective was to maximize total flow using two equivalent routes. In the present work, the objective is maximize the total flow past *Det3* of all incoming vehicles, mainline and on-ramp, not necessarily the total number of on-ramp vehicles that merge and travel beyond D. However, reducing transit times for the on-ramp vehicles is sometimes an additional benefit.

In Fig. 8a, the average travel time is shown for larger distance between merge regions, $L_{merge}$ = 3000 m, which makes the delay time for feedback longer. The cumulative number of merges is shown in Fig. 8b. In one hour, 540 merged at U and 408 at D which produced a total throughput at *Det3* of 3936 vehicles. The gaps in the D data especially are larger and the time between resets due to an empty D route is larger because of the increased delay. Nonetheless, nothing has been found to suggest that the system is going unstable.

A difference for $L_{merge}$ = 3 km compared to 1 km (Fig. 7) is the appearance of a jam moving upstream. As can be seen in the top panel of Fig. 9, a jam nucleated at $t$ = 2900 s on the mainline between the U and D merge regions. The jam propagates upstream past U in subsequent panels. It must eventually dissolve because the incoming flow of 1700/h/lane is not large enough to support a wide moving jam [9]. The larger distance between merge regions permits a jam to form in the pinch region near D before it enters the lower vehicle density upstream of U.

Using average velocity of vehicles on each route, as suggested by Lee *et al.* to improve estimates of average travel time and to obviate delay effects, does not appear to be worthwhile in the present scenario because so much time is spent in the congested regions near the on-ramps. Even if vehicles outside of congested areas traveled exceedingly fast, making the average velocity large, the delay would be substantially the same.

Simulations (not shown) for more than one hour showed that $T_{path}$ for both routes continued to fluctuate, but the trend to longer travel times diminished. The last ten minutes of the simulation shown in Fig. 4 are typical of the behavior after congested flow between the on-ramps has stabilized. A quasi-steady state, but not an equilibrium, was established for flow and average travel times.

3.2 *Controlled diversion*

Controlled diversion by means of directional signals is an alternative to driver choice based on average travel time. Following the methodology of Kerner's control algorithm ANCONA [9-11], in the present simulation the average mainline velocity $v_{ave}$ just upstream of the merge region at *Det1* (200 m beyond the beginning of the merge region) is compared to a preset congestion velocity, $v_{cong}$. That is, when $v_{ave} < v_{cong}$ a control action is taken. In ANCONA, a stop light is used to restrict the merge rate to a low value so that the mainline velocity returns to normal. In other words, synchronous flow is allowed to set in, but as soon as it does, the rate of merging from the on-ramp is restricted.

In the present work, instead of reducing the merge rate directly, on-ramp vehicles are diverted to a downstream ramp using directional signals when $v_{ave} < v_{cong}$. If the rate of diversion is large enough that the mainline velocity at the downstream ramp drops below $v_{cong}$, diversion is no longer allowed. Referring to Fig. 3, average velocities are measured in the right lane at detectors *Det1* and *Det2*. Averages are updated according to

$$v_{ave} = \alpha v + (1-\alpha)v_{ave}^{prev} \qquad (3)$$

whenever a new vehicle passes the detector with velocity *v*. Initially the average is set to the speed limit. The parameter *α* is taken to be 0.1. Total flow past the on-ramps is measured at *Det3*.

For the incoming flow considered in the Sec. 3.1 (1700 vehicles/h in each mainline lane and demand approaching the on-ramp U of 1000 vehicles/h), the cumulative number of merges for each ramp is shown in Fig. 10. At U, there were 672 merges and at D there were 322 with a total of 4174 vehicles passing *Det3* during one hour. In comparison to Fig. 4, the proportion of merges at U relative to D is more favorable, which results in higher total flow. The combined number of merges is 961 vehicles for one hour in Fig. 4 and a slightly higher 994 in Fig. 10.

The velocity near D is shown in Fig. 11. It oscillates regularly about $v_{cong}$ = 20 m/s with an average period of roughly 450 s. The oscillations develop large amplitude and the period increases with time. The rate of merging at D, Fig. 10, also shows the same oscillations. This behavior is similar to that found by Wahle *et al.* [2], but under different circumstances. The information that determines whether a vehicle is diverted from U to D is not based on travel time. Rather the dynamics are determined by the mainline response times of the average velocity to cessation and resumption of merging. The time for the diversion route to clear is also a factor.

The velocity as a function of position for vehicles at $t$ = 3600 s is shown in Fig. 12. The extent of synchronous flow is about 3 km from U, similar to that in Fig. 7. No wide jams formed during one hour, but vehicles were stopped briefly at $x$ = -2000 m. A similar fluctuation to near-zero speed was observed in Fig. 7b. Nearly periodic oscillations in velocity can be seen in the position-time plot of velocity on the mainline (averaged over both lanes) shown in Fig. 13.

To show how the behavior changes with increased separation of the on-ramps, results for $L_{merge}$ = 3 km are presented in Fig. 14. The cumulative number of merges is shown in Fig. 14a. After one hour, 669 vehicles merged at U and 298 at D. The total number of vehicles that passed *Det3*, measured at $x$ = 3100 m, was 4104, comparable to that in Fig. 10 and larger than the 3936 vehicles for driver choice (Fig. 8). The average period of oscillations in velocity at *Det2* (located at $x$ = 2500 m), as shown in Fig. 14b, is 490 s, which is somewhat larger than the average of 450 s for $L_{merge}$ = 1 km (Fig. 11). Interestingly, the period for infinite $L_{merge}$ was found to be less than 400 s for the same value of $v_{cong}$. Diversion to infinity represents the pure response time of the mainline velocity to the cessation and resumption of merging without the influence of downstream congestion. Wahle *et al.* [2] found that the cycle time was proportional to the length of the route, but in the examples analyzed in the present work, response time is the largest factor.

The velocity of vehicles at *t* = 3600s versus position is depicted in Fig. 14c. The region of synchronous flow upstream of U is similar to that in Figs. 12 and 13. Thus, the method of controlled diversion using average velocities just upstream of the ramps is also effective at the larger distance between U and D.

The period of velocity oscillations depends on $L_{merge}$ in addition to the response time of mainline flow to perturbations caused by changes in the rate of merging. In Fig. 15, the time T of the first peak in the autocorrelation function for different distances between on-ramps is shown. The velocity data of 90-minute simulations was used with the first 1800 s omitted because the period tends to lengthen as congestion is built up. The values for the first peak were found, therefore, to be larger than the average periods reported above but the trend with increasing $L_{merge}$ is similar.

The final example considered is for larger incoming flow, 2100 vehicles/h/lane, which is well above the flow required to sustain a wide moving jam. The velocity near D (*x* = 2500 m) is shown as a function of time in Fig. 16. The average period is 575 s, somewhat longer than for an incoming flow of 1700/h/lane (Fig.14b). In a two-hour interval, 740 vehicles merged at U and 457 at D, essentially all of the on-ramp demand during this time. The oscillating pattern remained stable over this extended period of time.

Velocity versus position for *t* = 2400, 2500 … 3600 s is shown in Fig. 17. A jam formed between *t* = 2300 and 2400 s in the region -3 to -2 km, which is in the pinch region associated with U [9]. The average velocities at both detectors were below $v_{cong}$ at this time. Because of the large incoming flow, the jam grew and propagated upstream. No jam nucleated prior to $t = 2300$ s. Controlled diversion allowed too many vehicles to merge to prevent nucleation of a jam. If the vehicles had been diverted to infinity, only about 315 would have merged at U in the first hour and a jam would not have been formed.

Once a wide moving jam forms, the mainline flow downstream of the front of the jam is limited to $q_{out} = 1810$/h/lane. In Fig. 18, the cumulative flow of both lanes at $x = -500$ m with a steady flow of 3672/h subtracted out is shown. The flow clearly levels off at $\sim 2q_{out}$ [9].

To assess the penalty for on-ramp drivers under the second scheme where they do not optimize individual travel times, $T_{path}^{(U)}$ and $T_{path}^{(D)}$ are plotted as a function of time in Fig. 19. Also shown is the number of vehicles on the diversion route at each time. The length of the route is 5400 m since $L_{merge} = 5000$ m. Typically the difference in travel times for the routes is no more than 2.5 minutes for trips lasting 5 to 8 minutes during the latter portion of the simulated time. Because the travel time information drivers use to choose their routes in Sec. 3.1 is delayed, the differences in actual travel times (Figs. 4 and 8a) are similar to those in Fig. 19. Thus the penalty for the controlled diversion is not substantially larger.

A more revealing comparison is shown in Fig. 20. In this figure, the average travel times for controlled diversion and for driver choice are plotted for the two routes, U and D, in a situation with large delayed feedback, $L_{merge} = 5000$ m. For each route, controlled diversion produces shorter travel times for the overwhelming majority of on-ramp vehicles.

## 4. Conclusions

The analysis presented in this paper of a double on-ramp configuration contains aspects of decision dynamics as regards route choice and control of merging rate by diversion. As previously explained by Kerner and Klenov [16], mainline congestion at a downstream on-ramp can impact the flow at an upstream on-ramp. The question addressed in the present work was how to apportion a given demand (incoming flow) between two on-ramps to achieve optimal results. Specifically, some of the vehicles attempting to merge were diverted to a lower-speed parallel road where they could merge at a downstream on-ramp. Two separate schemes were considered. In the first, drivers could choose one of two routes based on the average travel times. One route was to merge first and then proceed on the mainline and the other route was to divert to the parallel road and merge downstream. In the second scheme, vehicles were directed to merge or to divert based on the average mainline velocity just upstream of the ramp. If this velocity dropped below a preset value (20 m/s) at the upstream ramp, then vehicles were diverted unless the average velocity at the downstream ramp was also low. In neither scheme were vehicles prohibited from merging at both ramps at the same time. No metering was involved, just diversion.

Previous work by Wahle *et al*. [2] had shown that large oscillations in flow on equivalent routes develop when drivers choose on the basis of travel times. They attributed this behavior to delayed feedback because the travel time information was no more recent than the last vehicle to complete the route. Therefore, it was not surprising that related behavior was observed for the first scheme, even though the alternative routes were inequivalent. The near periodicity in flow and number of vehicles on each route found by Wahle *et al*. was replaced in the present situation by irregular oscillations. After about an hour, a quasi-steady state but not an equilibrium was found for flow and travel times. Since a large portion of the travel time was spent in the congested regions near ramps, using the instantaneous average velocity of all vehicles on the route as suggested by Lee *et al*. [3] was not useful.

The driver-choice scheme diverted sufficient numbers of vehicles to the downstream on-ramp to obtain reasonable throughput (total flow beyond the downstream merge area). However, it did not prevent jam formation for large demand and large separation between ramps. Such jams nucleated in the pinch region between the on-ramps and propagated upstream. They eventually dissolved if the upstream mainline flow was less than $q_{out}$ ≈1810 vehicles/h/lane, the incoming flow needed to sustain a jam [9]. Otherwise, a wide moving jam was produced.

In the second scheme, vehicles were directed to divert based upon average mainline velocities near the on-ramps. In this scheme the reduction in the rate of merging required to limit congestion is accomplished (to the extent possible) by diverting demand, not limiting it by metering with the potential formation of long queues. The pattern of merging and the number of vehicles on the diversion route showed regular, nearly periodic oscillations reminiscent of the equivalent route model analyzed by Wahle *et al*. [2]. However, unlike that model the period of oscillations in the present work was dominated by the response time of the mainline flow to interruption of merging, not the delayed feedback associated with travel times. There was a component linear in route length (and thus representative of delayed feedback), but it was not the dominate factor. Even if the vehicles were diverted to infinity, the velocity and merge rate at the upstream on-ramp was periodic.

Comparison of the two schemes revealed that throughput was larger with controlled diversion and the travel times for merging vehicles less than when on-ramp drivers attempted to optimize their own driving times. Controlled diversion permitted slightly higher merge rates with larger throughput (Fig. 21) by apportioning the demand more optimally between the ramps (larger merge rate upstream). Nonetheless, if demand and mainline flow were large enough, congestion with jam formation could not be avoided because neither scheme restricted flow severely enough.

**Figures**

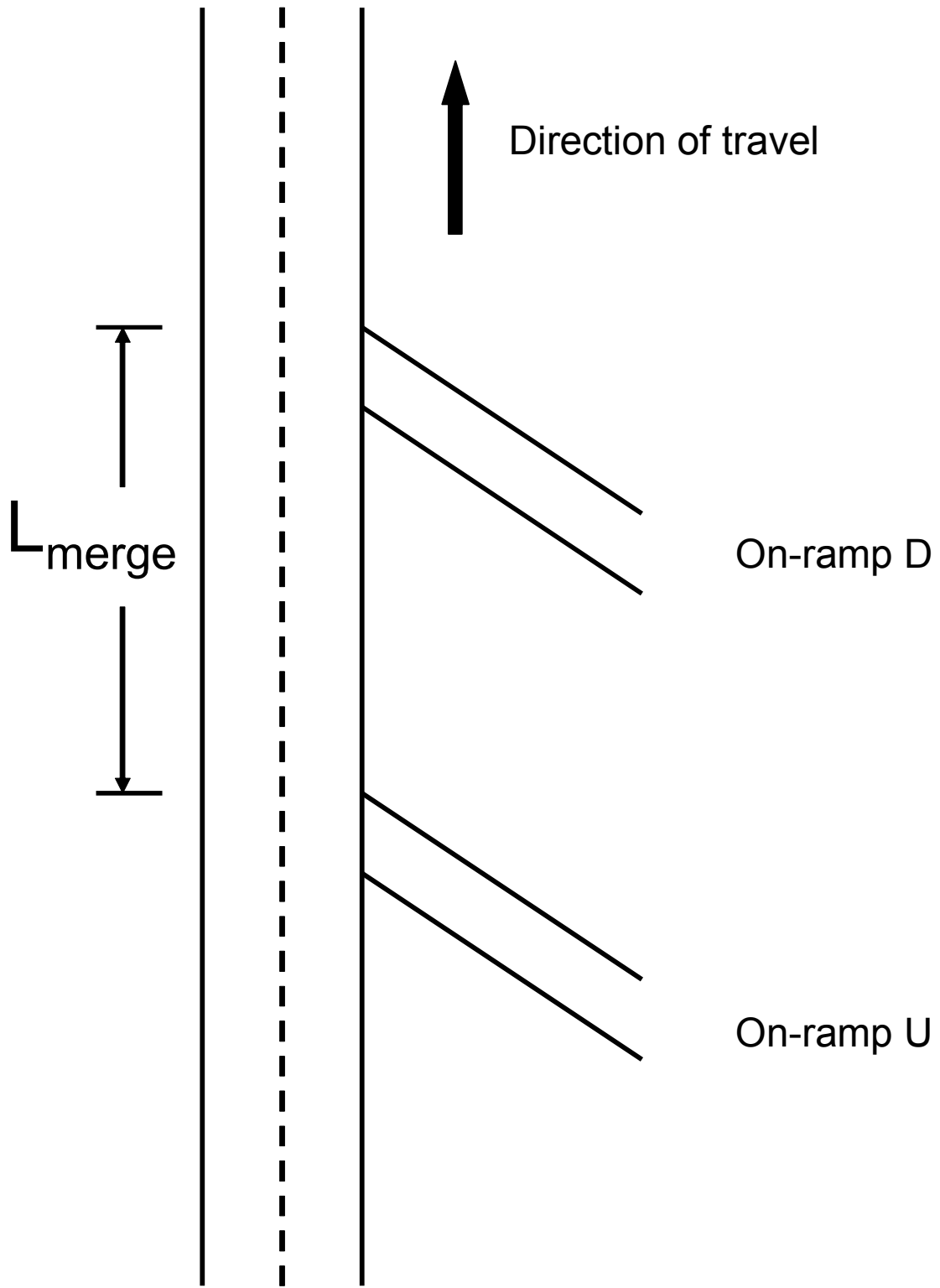

Fig. 1. Two-lane freeway with two on-ramps separated by distance $L_{merge}$.

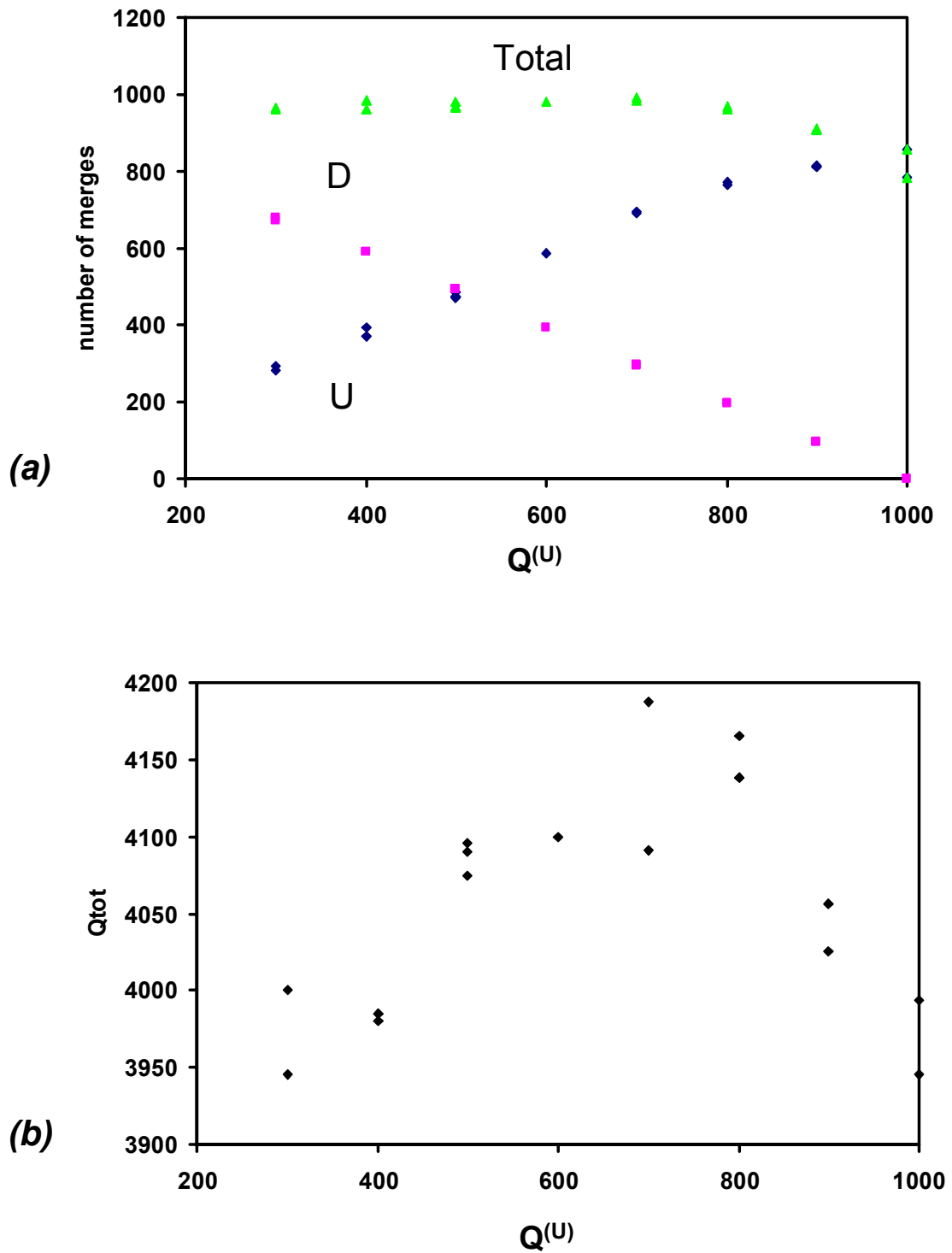

Fig. 2. (a) The number of merges in one hour at U, diamonds; at D, squares; and the total, triangles; as a function of $Q^{(U)}$ where $Q^{(U)} + Q^{(D)} = 1000$ vehicles in one hour. $Q^{(U)}$ and $Q^{(D)}$ are the demand (incoming on-ramp flow) in one hour at U and D. (b) The total mainline flow $Q_{tot}$ (for both lanes) just beyond D. Incoming flow is 1700/h/lane and $L_{merge} = 1000$ m.

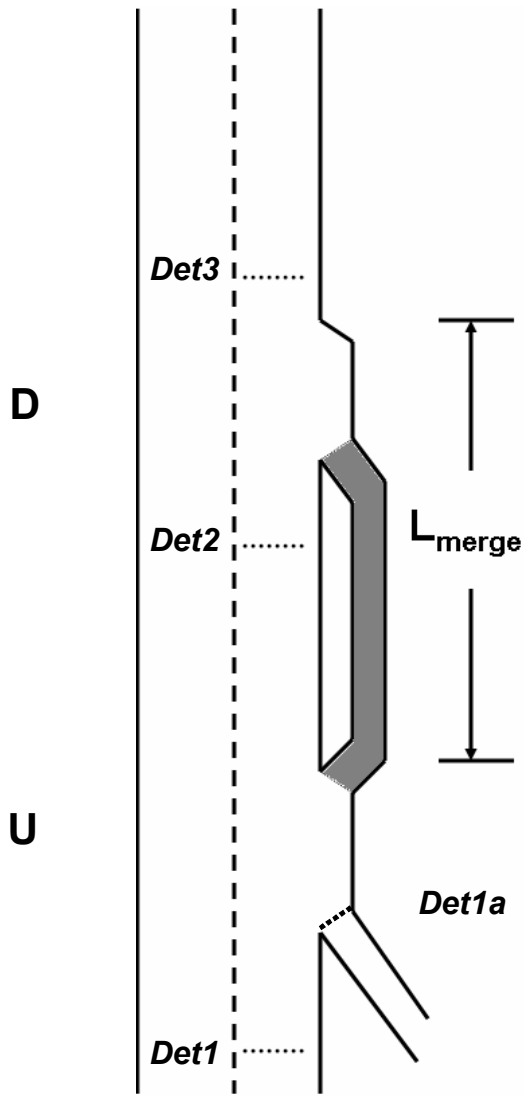

Fig. 3. Two-lane freeway with on-ramp and diversion route (shaded). Detectors *Det1* and *Det2* are located upstream of the merge regions (200 m from beginning) and Det3 is located 100 m down from end of downstream merge region. The length of the merge regions is 300 m. *Det1a* is located on the on-ramp at the beginning of the merge region.

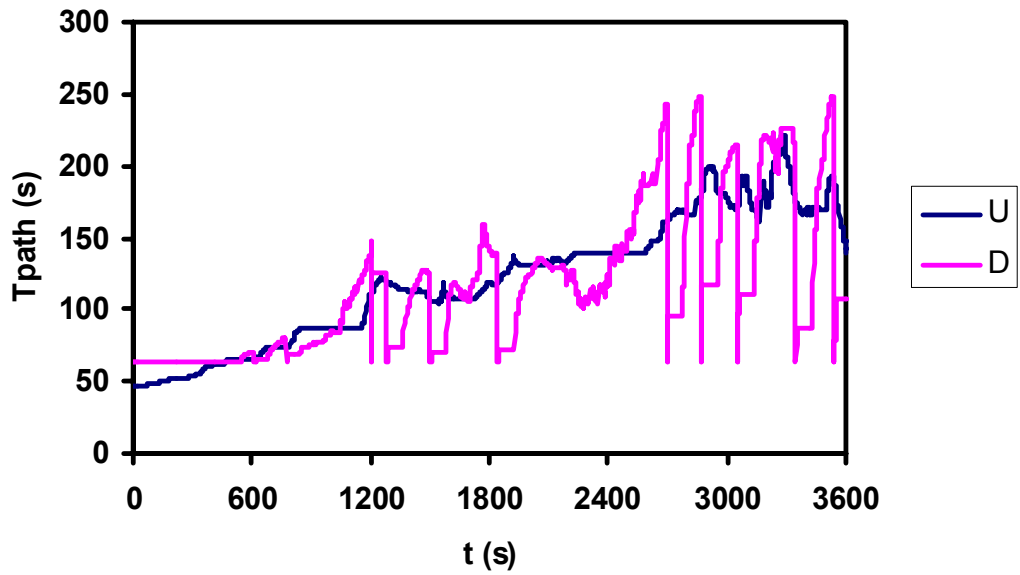

Fig. 4. Average travel time [Eq. (1)] for two routes as a function of time. The route labeled U involves merging at the upstream on-ramp and the route labeled D involves diversion followed by merging at the downstream merge area. The time to travel from detector *Det1a* to *Det3* (see Fig. 3) via either path determines the travel time. The incoming flow on the mainline is 1700/h/lane and on-ramp demand is 1000/h. $L_{merge}$ is 1 km.

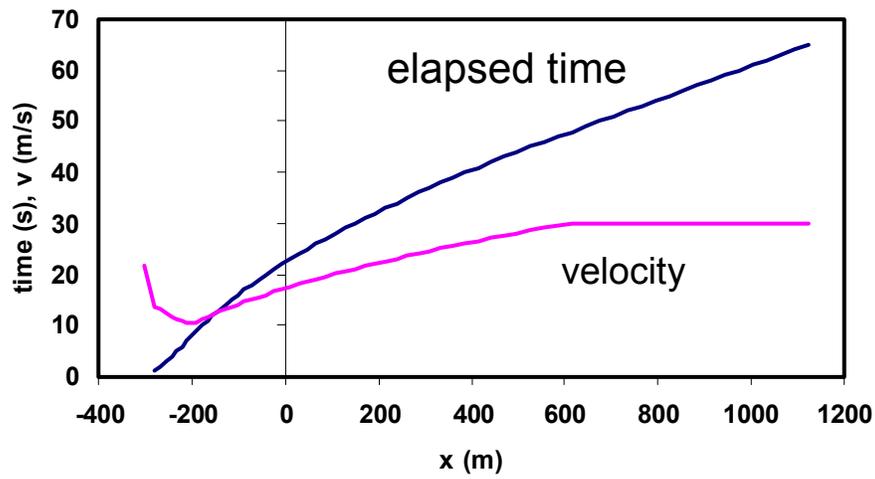

Fig. 5. The elapsed time and velocity of the 100[th] vehicle to merge at U as a function of vehicle position. The vehicle merged just as it entered the merge region.

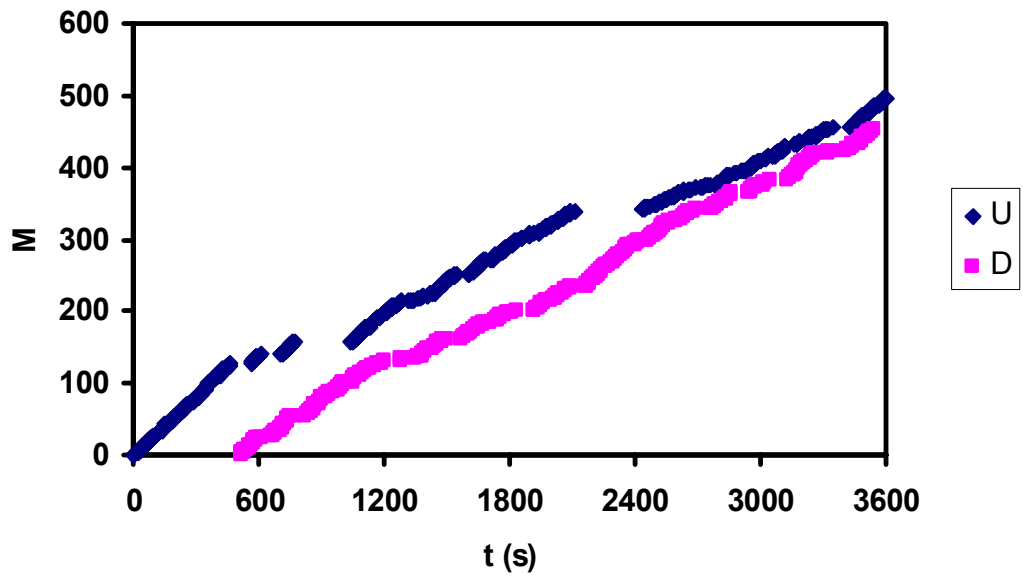

Fig. 6. The cumulative number of merges at the upstream merge region U and the downstream merge region D as a function of time associated with Fig. 4.

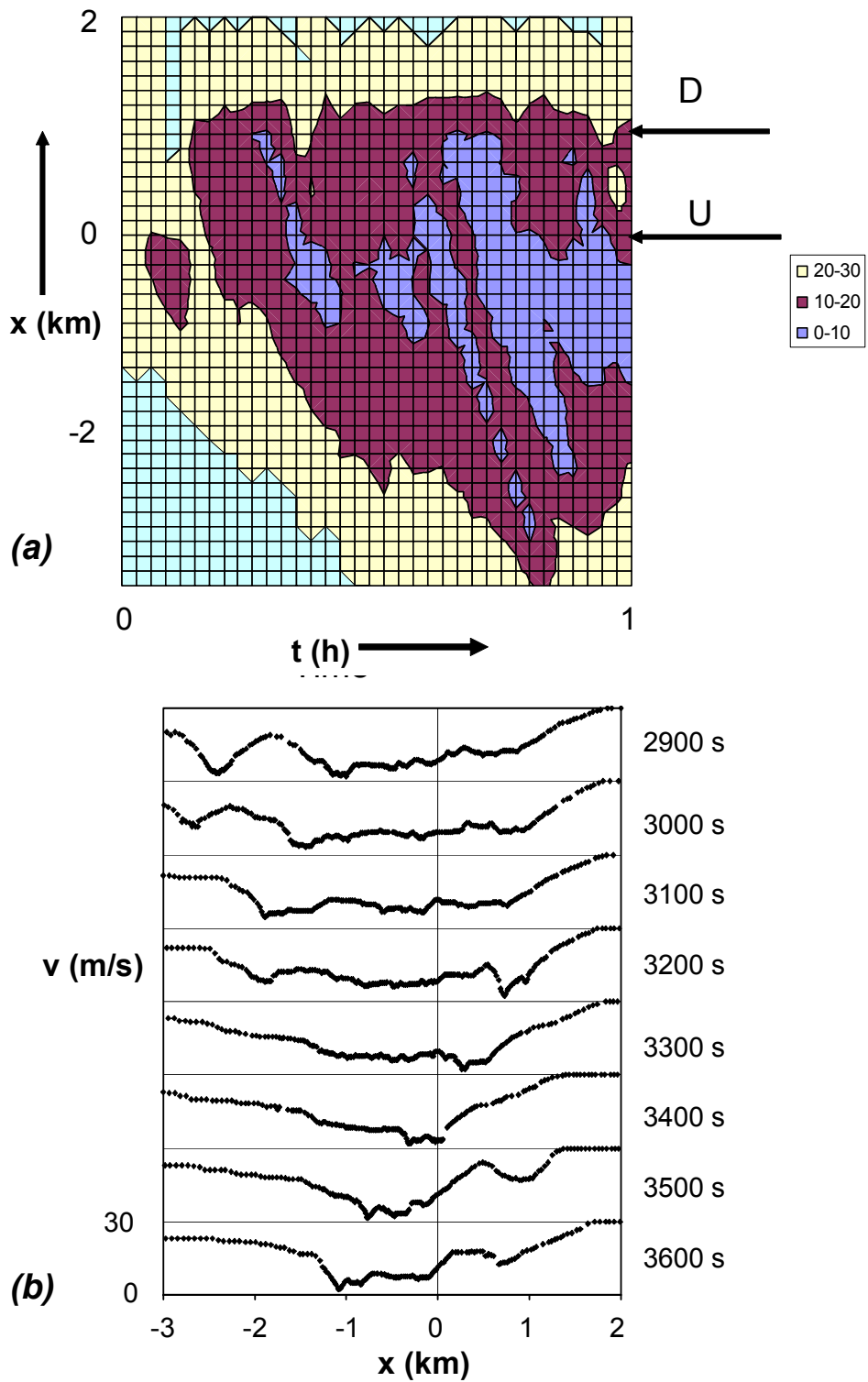

Fig. 7. (a) A map of mainline velocity (both lanes) in position-time. The distance spanned is five km and the time elapsed is one hour. The two merging regions are marked with arrows U and D. They are separated by 1 km. (b) Velocity of vehicles in the right lane of the mainline for the last 800 s. Each panel is offset by 30 m/s. The top panel is at $t = 2900$s and subsequent panels are at 100-s intervals.

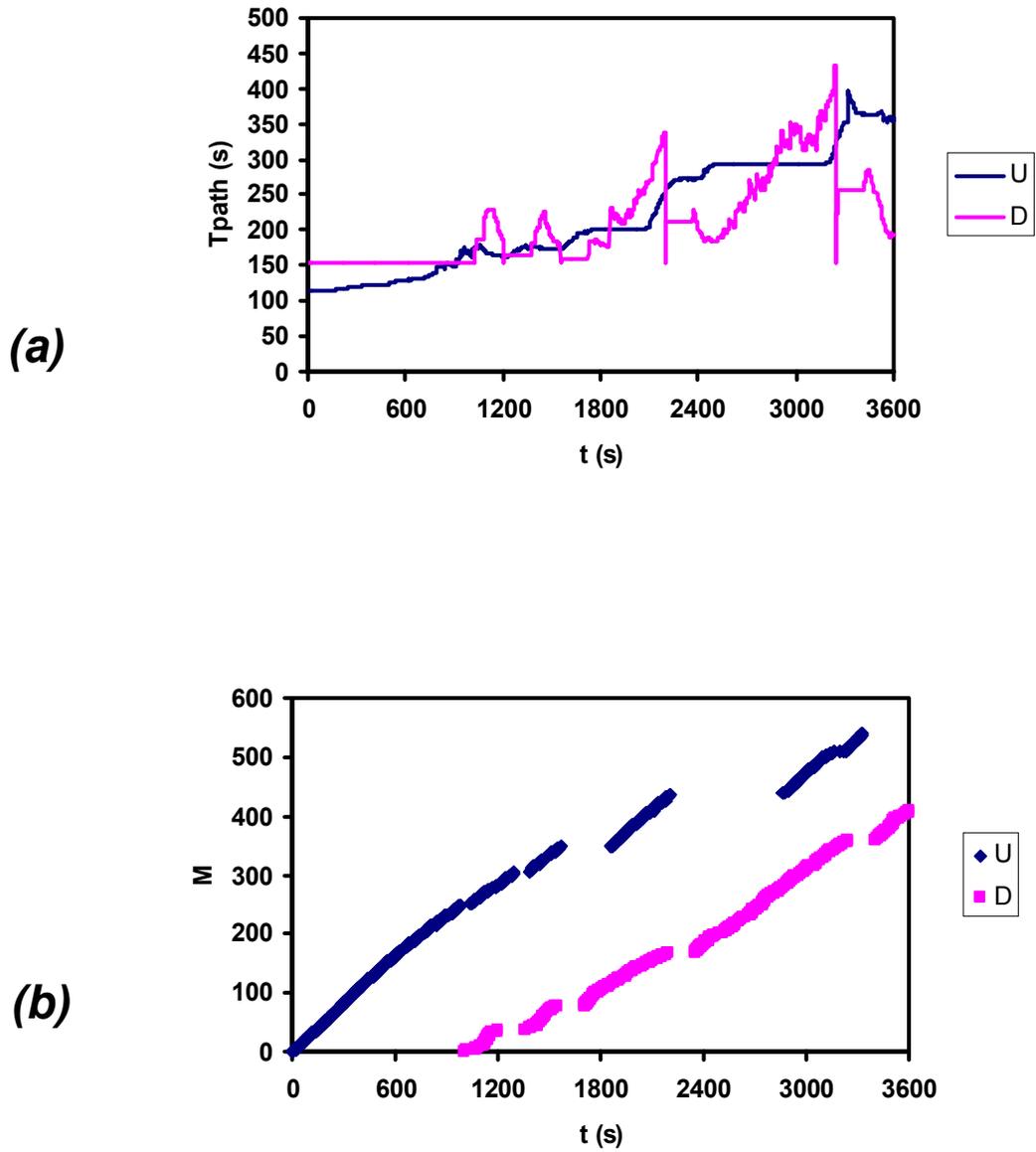

Fig. 8. (a) Average travel time for U and D paths. (b) Cumulative number of merges as a function of time. $L_{merge}$ = 3 km. Same incoming flow and demand as in Figs. 4-7.

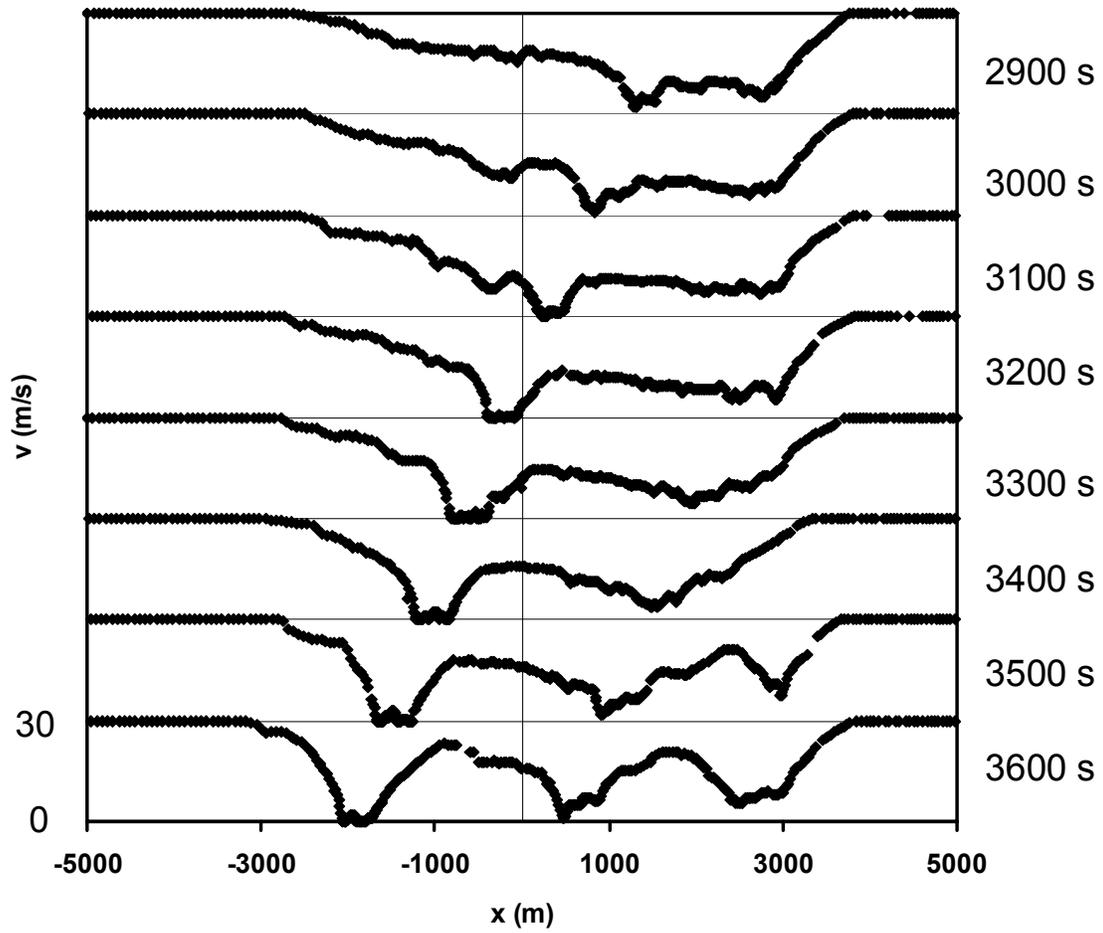

Fig. 9. Velocity in the right lane of the freeway as a function of distance at 100-s intervals of time. Each curve is offset vertically by 30 m/s. The lowest panel is for $t = 1$ h. Parameters are the same as in Fig. 8.

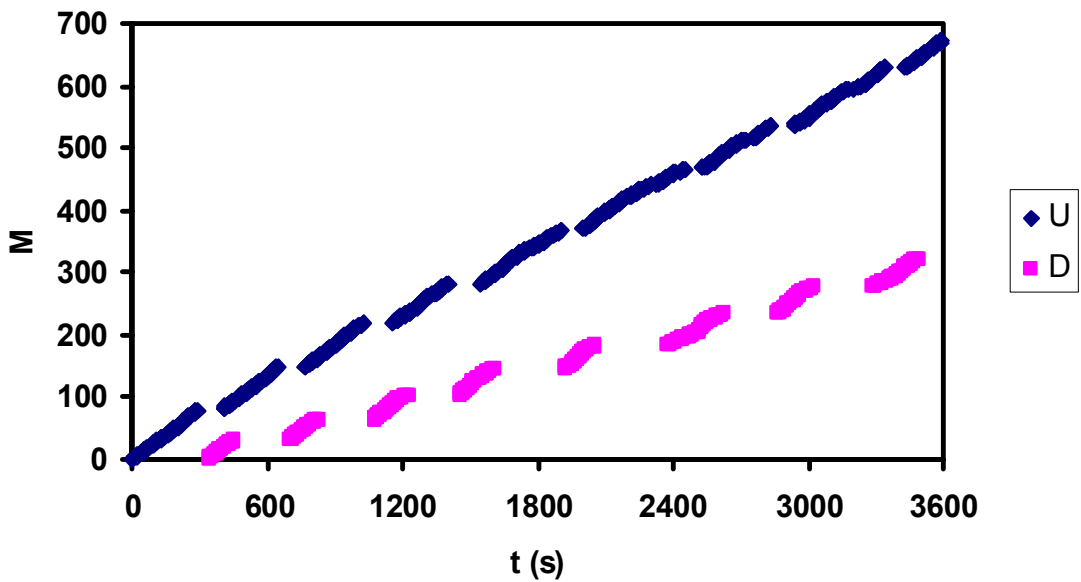

Fig. 10. Cumulative number of merges at the upstream (U) and downstream (D) on-ramps as a function of time. The distance between merge regions is $L_{merge}$ = 1 km. The critical velocity to trigger diversion is $v_{cong}$ = 20 m/s. The incoming flow on the mainline is 1700/h/lane and the on-ramp demand is 1000/h.

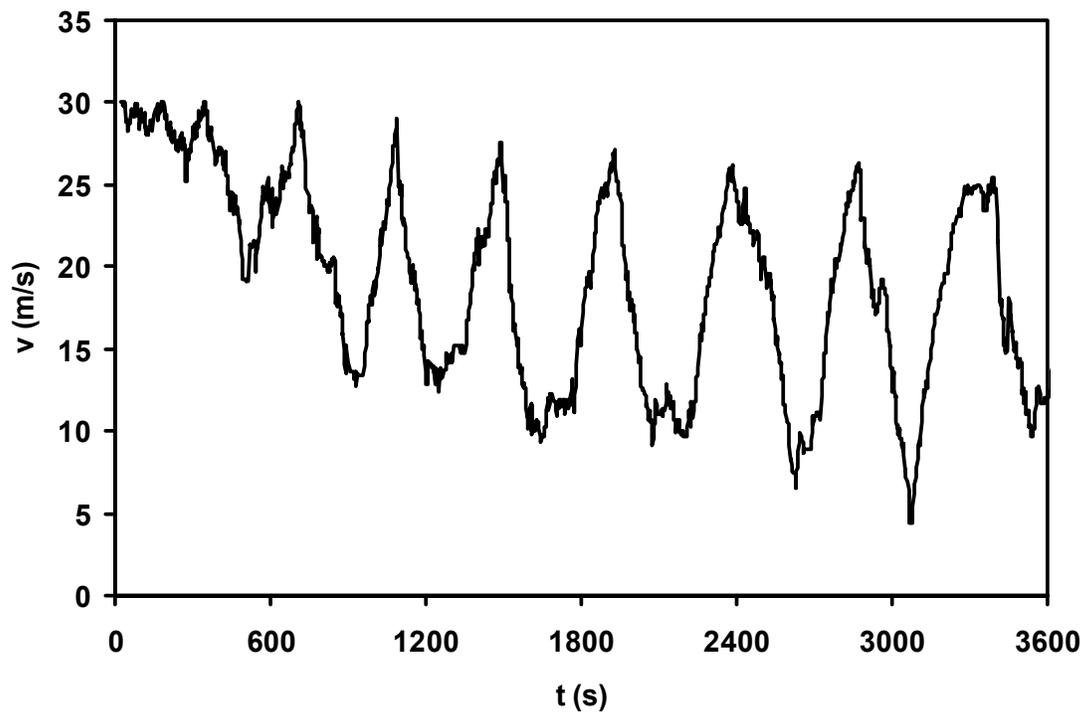

Fig. 11. The velocity at *Det2* near the downstream on-ramp as a function of time. Parameters are the same as Fig. 10.

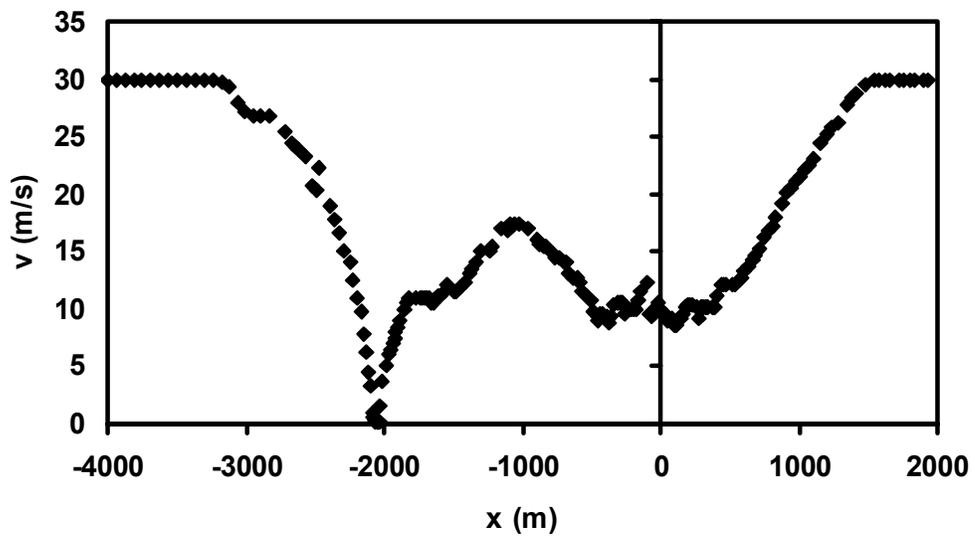

Fig. 12. Velocity versus position for the right lane of the mainline at $t = 3600$ s. Control is based on the average mainline velocities at *Det1* and *Det2* located at -500 m and 500 m, where comparison to $v_{cong} = 20$ m/s is made. Parameters are the same as for Fig. 10.

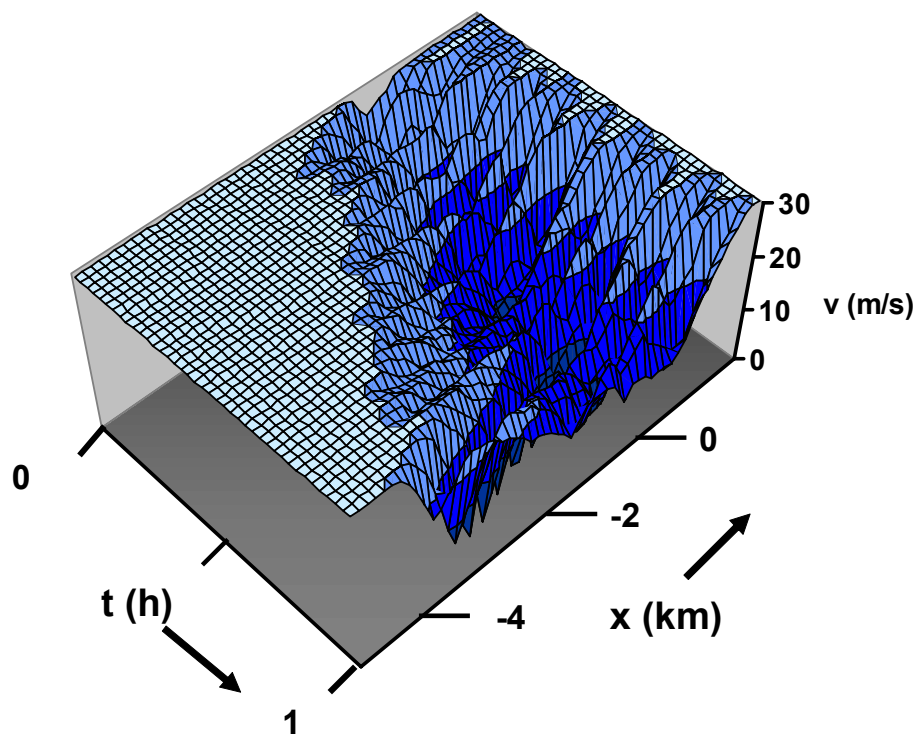

Fig. 13. Velocity in position-time for one hour elapsed time. On-ramps are at $x = 0$ (U) and $x = 1$ km (D). Parameters same as for Fig. 10. Both lanes are included.

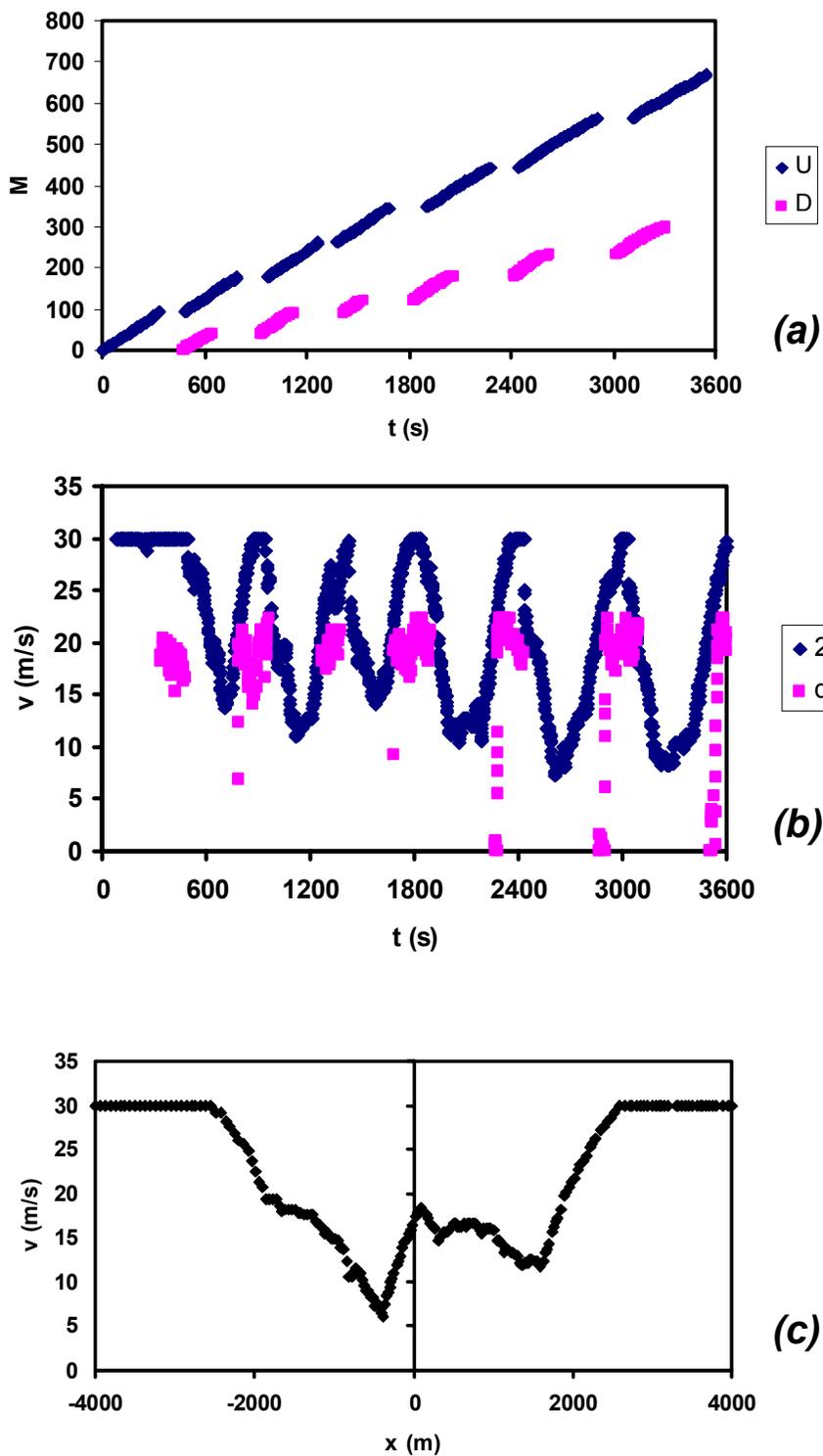

Fig. 14. Results for $L_{merge}$ = 3000 m; other parameters same as Fig. 10. (a) Cumulative number of merges at U and D as a function of time. (b) Velocity on the mainline at $x$ = 2500 m. Data labeled "diverts" is the velocity of vehicles diverting at U. (c) Velocity in the right lane versus position at $t$ = 3600 s.

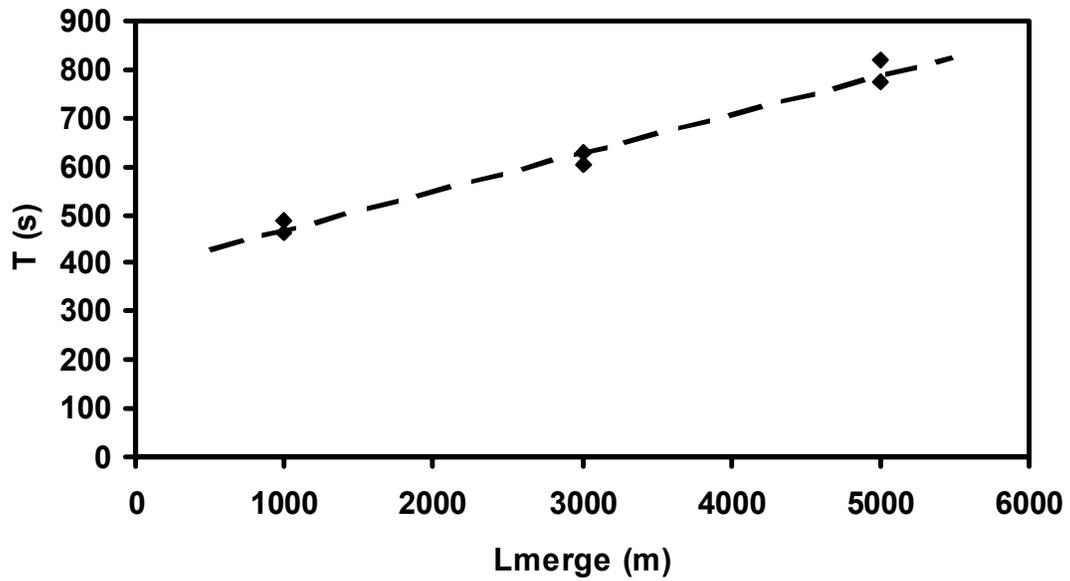

Fig. 15. The time of the first peak in the autocorrelation function for the velocity at *Det2* versus $L_{merge}$. Data for $t > 1800$ s used from 90-minute simulations. The critical velocity to trigger diversion is $v_{cong} = 20$ m/s. The incoming flow on the mainline is 1700/h/lane and the on-ramp demand is 1000/h.

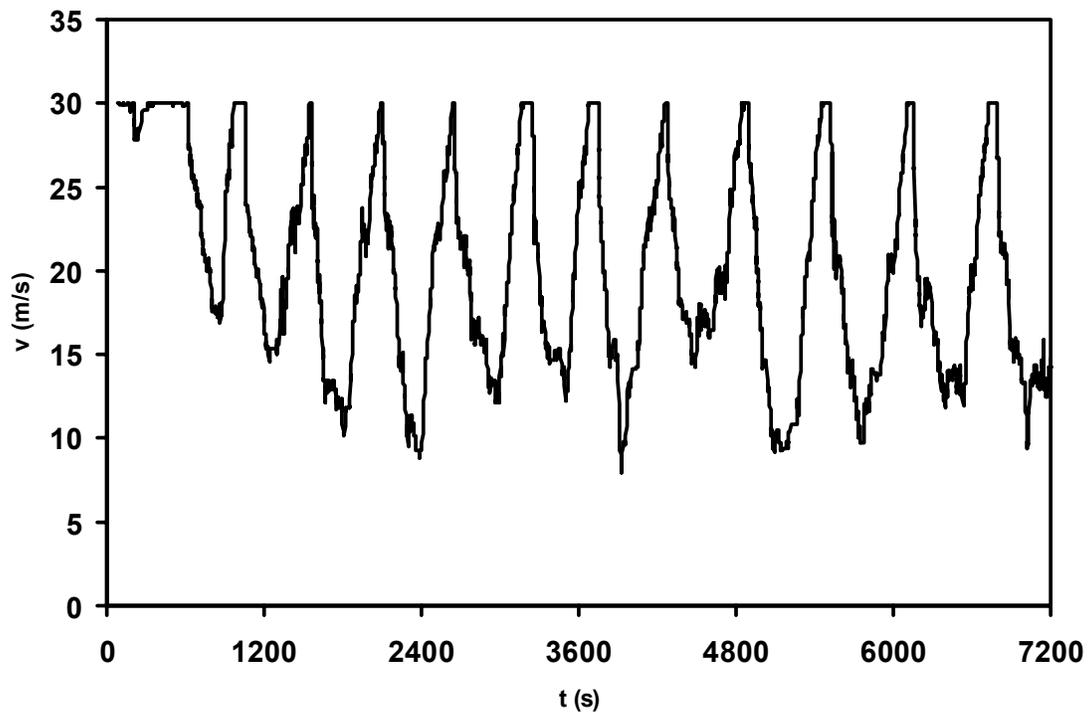

Fig. 16. The velocity as a function of time at *Det2*. The incoming flow is 2100 vehicles/h/lane and the on-ramp demand is 600/h. $L_{merge}$ = 3 km and $v_{cong}$ = 20 m/s.

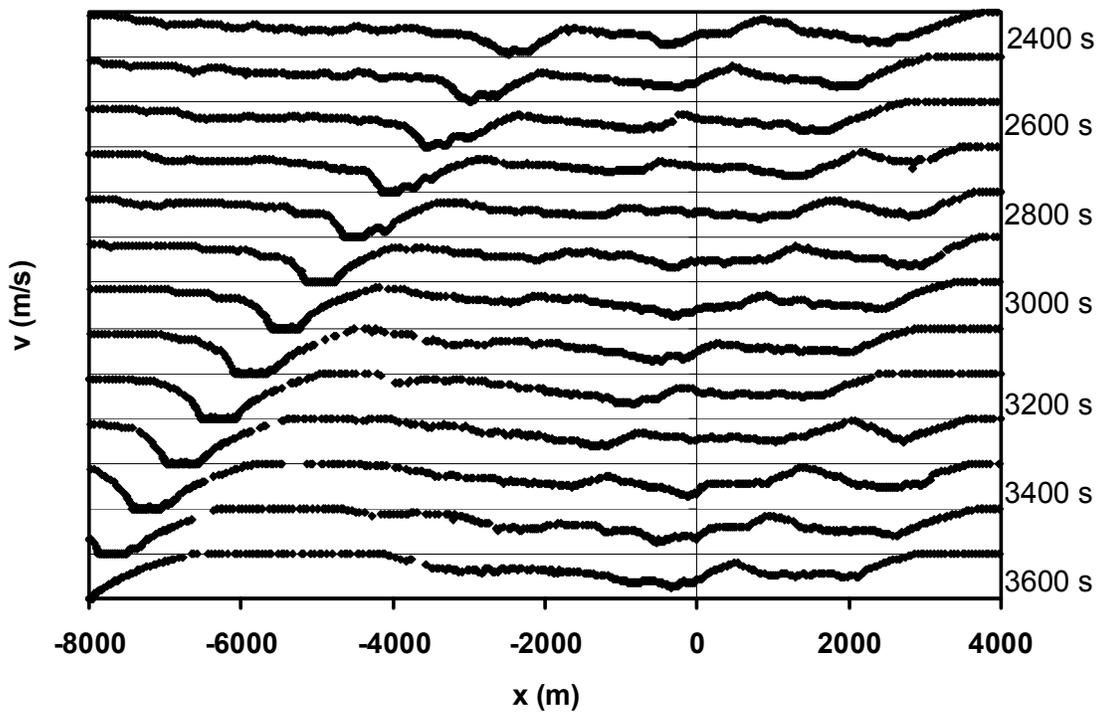

Fig. 17. Velocity versus position for $t = 2400$ to $3600$ s. Each panel differs by 100 s and is offset by 30 m/s. The downstream on-ramp is at $x = 3000$ m. Parameters are the same as for Fig. 16. The velocities of vehicles in the left lane, not shown, are nearly identical.

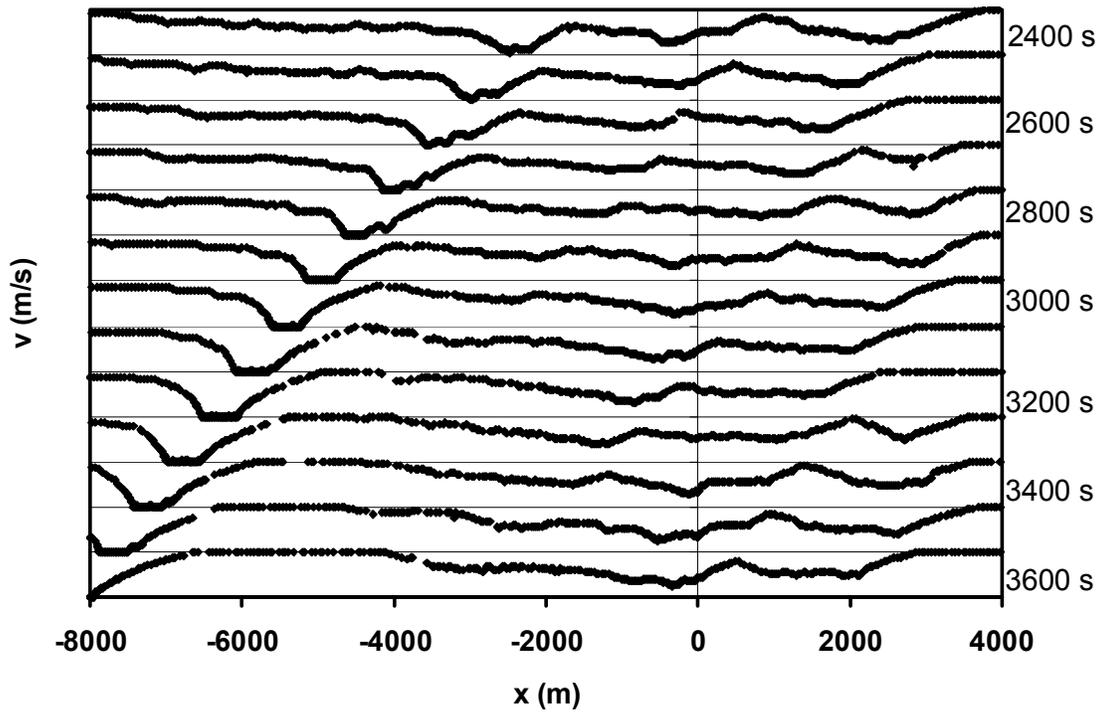

Fig. 18. The cumulative mainline flow of both lanes past $x = -500$ m in excess of 3672 vehicles/hour. The flow levels off at about $q_{out} = 1810$/h/lane after the formation of a wide moving jam at $t > 2300$ s as shown in Fig. 17.

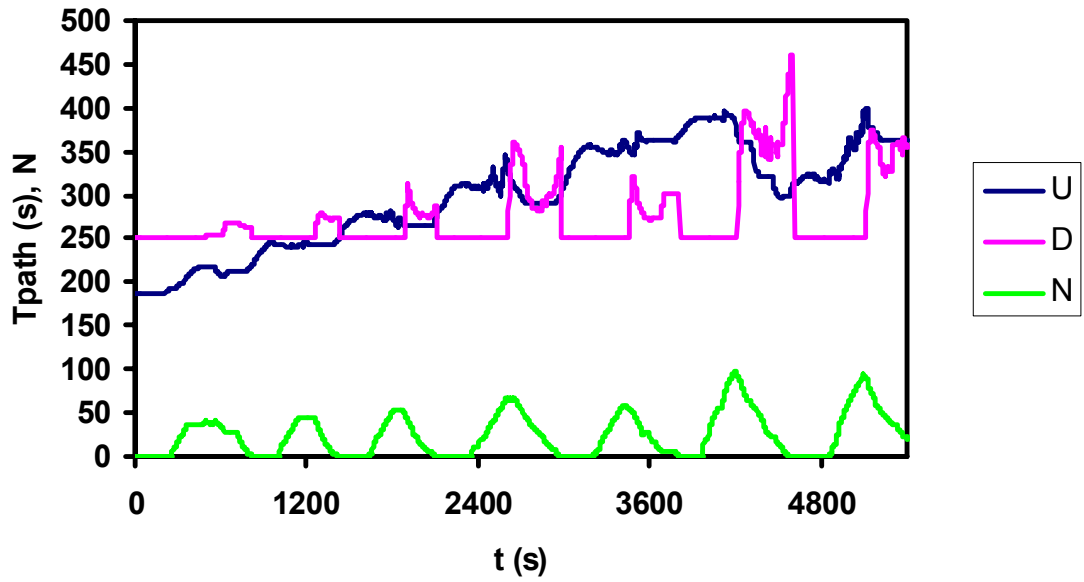

Fig. 19. Average travel time for the U and D paths as a function of time for $L_{merge}$ = 5000 m. Other parameters are the same as Fig. 10. The lowest curve is the number of vehicles $N$ on the diversion route.

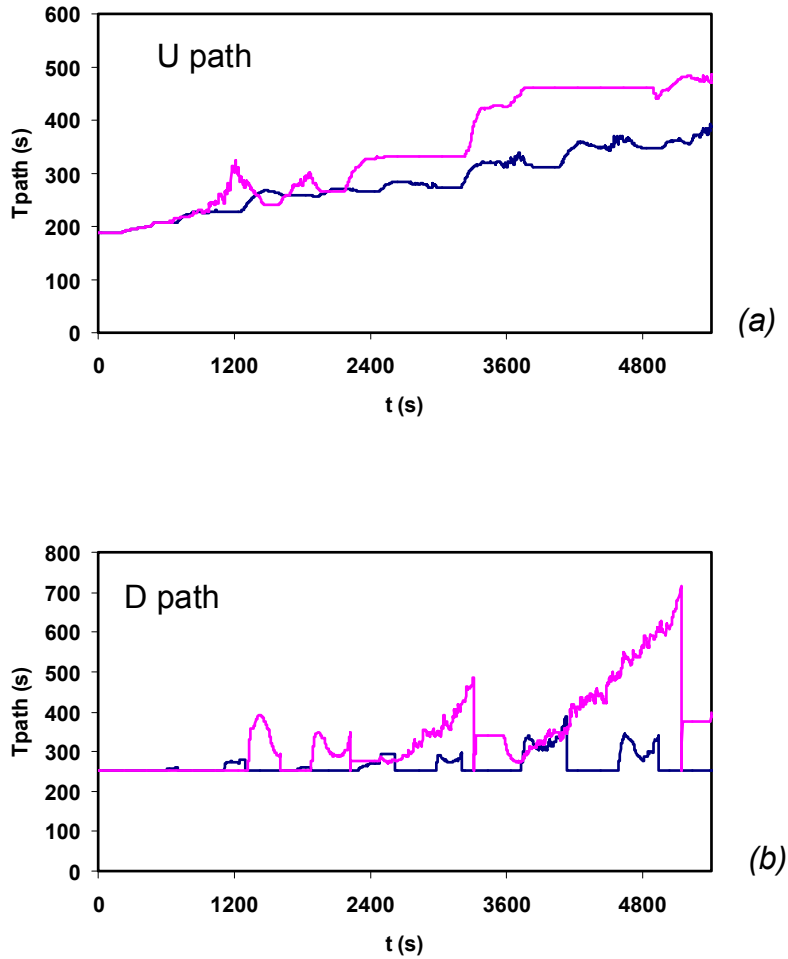

Fig. 20. The average travel times for controlled diversion (blue) are compared to those when drivers choose their routes (red) based on messages showing (a) $T_{path}^{(U)}$ and (b) $T_{path}^{(D)}$. The distance between merge regions is 5 km. The incoming flow on the mainline is 1700/h/lane and the on-ramp demand is 1000/h.

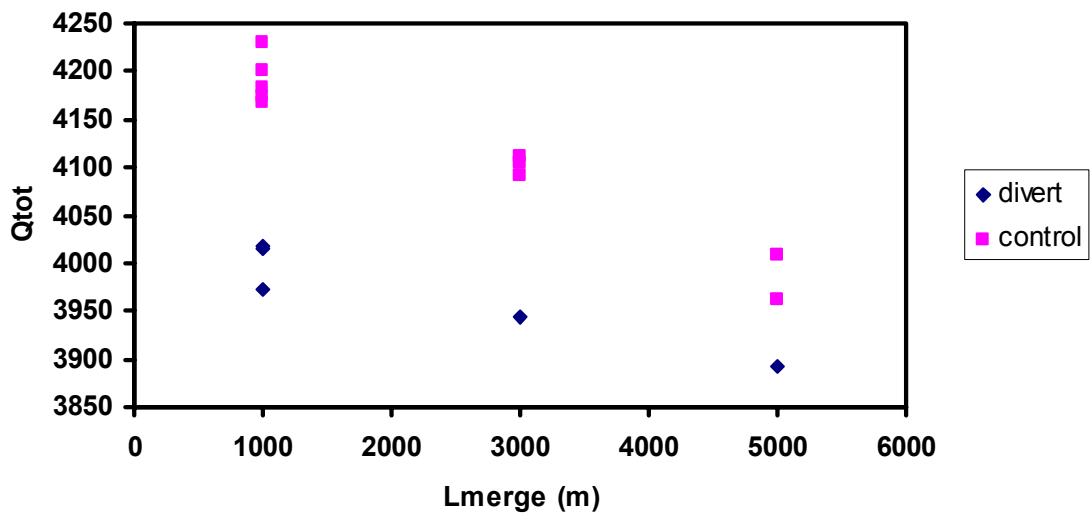

Fig. 21. The total throughput in one hour at *Det3* versus $L_{merge}$. The data labeled "divert" pertains to diversion based on average travel times ($T_{path}^{(U)}$ and $T_{path}^{(D)}$) and that labeled "control" is based on the average velocities on the mainline at *Det1* and *Det2*. The incoming flow on the mainline is 1700/h/lane and the on-ramp demand is 1000/h.